\documentclass[10pt,letterpaper]{article}
\usepackage[top=0.85in,left=2.75in,footskip=0.75in]{geometry}

\usepackage{amsmath,amssymb}

\usepackage{changepage}

\usepackage{textcomp,marvosym}

\usepackage{cite}

\usepackage{nameref,hyperref}

\usepackage[right]{lineno}

\usepackage[nopatch=eqnum]{microtype}
\DisableLigatures[f]{encoding = *, family = * }

\usepackage[table]{xcolor}

\usepackage{array}

\newcolumntype{+}{!{\vrule width 2pt}}

\newlength\savedwidth

\raggedright
\usepackage[aboveskip=1pt,labelfont=bf,labelsep=period,justification=raggedright,singlelinecheck=off]{caption}

\makeatletter
\renewcommand{\@biblabel}[1]{\quad#1.}
\makeatother

\usepackage{lastpage,fancyhdr,graphicx}
\usepackage{epstopdf}
\fancyheadoffset[L]{2.25in}
\fancyfootoffset[L]{2.25in}
\begin{document}
\vspace*{0.2in}

\begin{flushleft}
{\Large
\textbf\newline{Environmental variability promotes the evolution of cooperation among geographically dispersed groups on dynamic networks} 
}
\newline

Masaaki Inaba\textsuperscript{*1},
Eizo Akiyama\textsuperscript{2}

\bigskip
\textbf{1} Graduate School of Science and Technology, University of Tsukuba, 1-1-1 Tennodai, Tsukuba, Ibaraki, Japan

\textbf{2} Institute of Systems and Information Engineering, University of Tsukuba, 1-1-1 Tennodai, Tsukuba, Ibaraki, Japan
\bigskip

* masaaki.inaba@gmail.com
\end{flushleft}


\section*{Abstract}
The evolutionary process that led to the emergence of modern human behaviors during the Middle Stone Age in Africa remains enigmatic.
While various hypotheses have been proposed, we offer a new perspective that integrates the variability selection hypothesis (VSH) with the evolution of cooperation among human groups.
The VSH suggests that human adaptability to fluctuating environments was a primary force driving the development of key evolutionary traits.
However, the mechanisms by which environmental variability (EV) influenced human evolution, particularly the emergence of large-scale and complex cooperative behaviors, are not yet fully understood.
To explore the connection between intensified EV and the evolution of intergroup cooperation, we analyzed three stochastic models of EV:
(i) Regional Variability (RV), where resource-rich areas shift while overall resource levels remain stable;
(ii) Universal Variability (UV), where overall resource levels fluctuate but resource-rich areas remain stable; and
(iii) Combined Variability (CV), where both resource-rich areas shift and overall resource levels fluctuate.
Our results show that RV strongly promotes cooperation, while UV has a comparatively weaker effect.
Additionally, our findings indicate that the coevolution of cooperation and network structures is crucial for EVs to effectively promote cooperation.
This study proposes a novel causal link between EV and the evolution of cooperation, potentially setting a new direction for both theoretical and empirical research in this field.

\section*{Author summary}
This study investigates how changing environmental conditions may have influenced the emergence of cooperative behaviors among early human groups during the Middle Stone Age in Africa, a pivotal period in human evolution.
We present a novel approach to explaining the link between environmental changes and the evolution of cooperation by using models that simulate fluctuations in resource availability across various patterns.
Our findings suggest that cooperation is more likely to emerge and sustain when resource distribution across regions varies over time, in contrast to cases where overall wealth changes but remains concentrated in fixed areas.
Furthermore, our results indicate that the coevolution of cooperation and social structures is crucial in determining whether environmental changes foster cooperation, highlighting a dynamic interplay between environmental factors and social adaptability.
We believe these findings contribute to broader discussions in anthropology, archaeology, and the study of complex systems, enriching our understanding of human nature and society.

\section*{Introduction}
Deepening our understanding of the evolutionary origins of modern human behavior is essential for comprehending the nature of humanity and society.
In anthropology and archaeology, ``modern human behavior" refers to traits unique to or primarily associated with Homo sapiens, marked by abstract thinking, symbolic expression, complex planning, and ultrasociality.
These behaviors include language, religion, mythology, art, music, entertainment, humor, altruism, long-distance trade, and the creation of intergroup networks.
Numerous studies concur that these behavioral patterns emerged during the Middle Stone Age (MSA) in Africa\cite{Mcbrearty2000, Henshilwood2003, dErrico2020, Wilkins2021, Bergstrom2021}.
While there is broad consensus on when and where these behaviors originated, the mechanisms driving their emergence remain enigmatic, despite various proposed theories.

For several years, hypotheses \cite{Potts1996, Potts1998, Potts2013, Potts2018, Potts2020, TrauthMaslin2007, TrauthMaslin2010, TrauthMaslin2014, Ziegler2013, Kalan2020, Siepielski2017, Faith2021} attempting to explain the evolution of hominin behavior by focusing on environmental variability (EV) in Africa during the MSA have garnered significant attention.
Among these, Potts’ variability selection hypothesis (VSH) \cite{Potts1996, Potts1998} proposes that intensified environmental change favored ``versatilists" those capable of rapid adaptation to new environments over specialists, who adapt to specific environments, or generalists, who adapt across a range of environments.
Here, EV encompasses changes in landscape dynamics (such as land-lake oscillations), climate (like arid-moist climate oscillations), variations in flora and fauna, ultimately leading to the unpredictability of resource availability.
Initially, this hypothesis was supported by a temporal correlation between intensified environmental changes, the replacement of human species, and the increased complexity of cultural artifacts, such as stone tools and ornaments \cite{Faith2021}.
In addition, the cognitive buffer hypothesis (CBH) \cite{Schuck-Paim2008, Sol2008, Sol2009} provides a neuroscientific basis for VSH, and a mathematical model \cite{Grove2011} demonstrates its theoretical feasibility.
The CBH posits that larger brain sizes in animals, including humans, evolved as a buffer against environmental variability, enhancing survival through improved problem-solving and learning abilities.
In contrast, several theories \cite{Navarrete2011, Will2021, Stibel2023} propose that EV and behavioral diversity do not necessarily drive human encephalization.
These theories emphasize the role of social contexts, as suggested by the social brain hypothesis (SBH) \cite{Whiten1988, Dunbar1998, Barrett2007, Grove2008, Knight2011, Hayes2014, Faith2021, Dunbar2024a}, and consider focus such as dietary influences \cite{DeCasien2017, Grabowski2023}.
The SBH argues that human intellectual abilities evolved in response to the selection pressures of complex social environments, which required the effective management of social relationships within and between groups.
Therefore, much remains unknown about the impact of EV on the evolution of cognitive and behavioral traits in hominins.

Our study suggests that VSH, typically explained through the CBH, may also be connected to the SBH, which is generally considered separate from both VSH and CBH.
While complex social environments encompass various factors, what uniquely characterizes human societies is the extensive and sophisticated cooperation observed, including intergroup cooperation and trade, which contrasts with the intragroup cooperation common in many animal societies.
These advanced social behaviors are central to modern human behavior, and understanding their origins requires focusing on social factors that extend beyond individual-level adaptations, such as those proposed in CBH.
Specifically, we demonstrate that EV fosters intergroup cooperation, which may have contributed to the development of complex social structures.

There are several points of concern when using the term ``group."
First, groups within the complex social environment described by the SBH are nested in a series of fractal-structured networks \cite{Bird2019, Dunbar2020, Dunbar2024a}.
As a result, when smaller groups ally and cooperate to form a larger group, whether this cooperation is viewed as intragroup cooperation within the larger group or intergroup cooperation among the smaller groups depends on the level of analysis.
For simplicity, we assume a certain level of grouping and analyze their intergroup cooperation, though this could alternatively be seen as intragroup cooperation from the perspective of a higher-level group.
Furthermore, while treating groups as units of adaptation is highly debated in evolutionary biology \cite{Smith1976, Okasha2001, Eldakar2011}, our focus here is on cultural evolution rather than biological evolution.
In this cultural context, we assume that a group has a degree of autonomy, treating individual relationships and nested group structures as a black box.
Here, autonomy suggests that the basic behavioral patterns for a group regarding which groups it cooperates with or does not are influenced by intergroup interactions and evolve over time.

In the study of the evolution of cooperation, many studies have been conducted within the framework of evolutionary game theory \cite{Axelrod1981, Nowak2006, Szabo2007, Zaggl2014, Perc2017, West2021}, though most assume a stable environment.
Only a limited number of studies consider environmental factors in the evolution of cooperation, and these, typically in biological or physical contexts, focus on aspects such as extrinsic population variability \cite{Brockhurst2007, Miller2015}, variability in game structure \cite{Gokhale2016, Stojkoski2021}, variability in the strength of selection \cite{Assaf2013}, the impact of EV on learning strategies \cite{Borg2012}, and resource pressure \cite{Pereda2017}.
However, these studies do not fully address our research objective of understanding how EVs influences the evolution of cooperation.

Our research thus investigates how the unpredictability of resource acquisition (EV) may drive the evolution of cooperation among geographically dispersed groups, with a focus on the origins of the social aspects that characterize modern human behavior.

\section*{Model}

We use an agent-based simulation model within the framework of evolutionary game theory to investigate how increasing EV influences the evolution of cooperation.
Given the limited availability of detailed data on EV and the spatial distribution of hominid groups during the MSA in Africa, our model adopts a highly abstracted approach, aiming to reveal the general effects of EV based on reasonable assumptions while excluding specific details.
The model operates as follows: several geographically separated regions, each with varying levels of resource accessibility, experience fluctuating resource availability over time (EV).
Each region hosts a single group (agent), and interactions between groups, such as resource exchanges (game) and behavioral pattern transmission (reformation), affect and adjust their relationships.

\subsection*{Agent and structure}

In this model, each agent represents a single group and has a strategy of either cooperation or defection (in this study, ``C" represents a cooperation strategy or a cooperator, while ``D" represents a defection strategy or a defector).
Initially, all agents are set to D, as intergroup cooperation is considered extremely unlikely \cite{DeDreu2022a, Rodrigues2023}.
This study focuses exclusively on intergroup interactions and does not consider intragroup interactions.
While a two-level model would be necessary to analyze the tension between intergroup and intragroup interactions, as seen in multilevel selection studies, a one-level model is suitable here given our focus on intergroup interactions.
All agents are suited within a geographic structure, forming an interaction structure.

\begin{figure}[ht]
    \centering
    \includegraphics[width=1.0\linewidth]{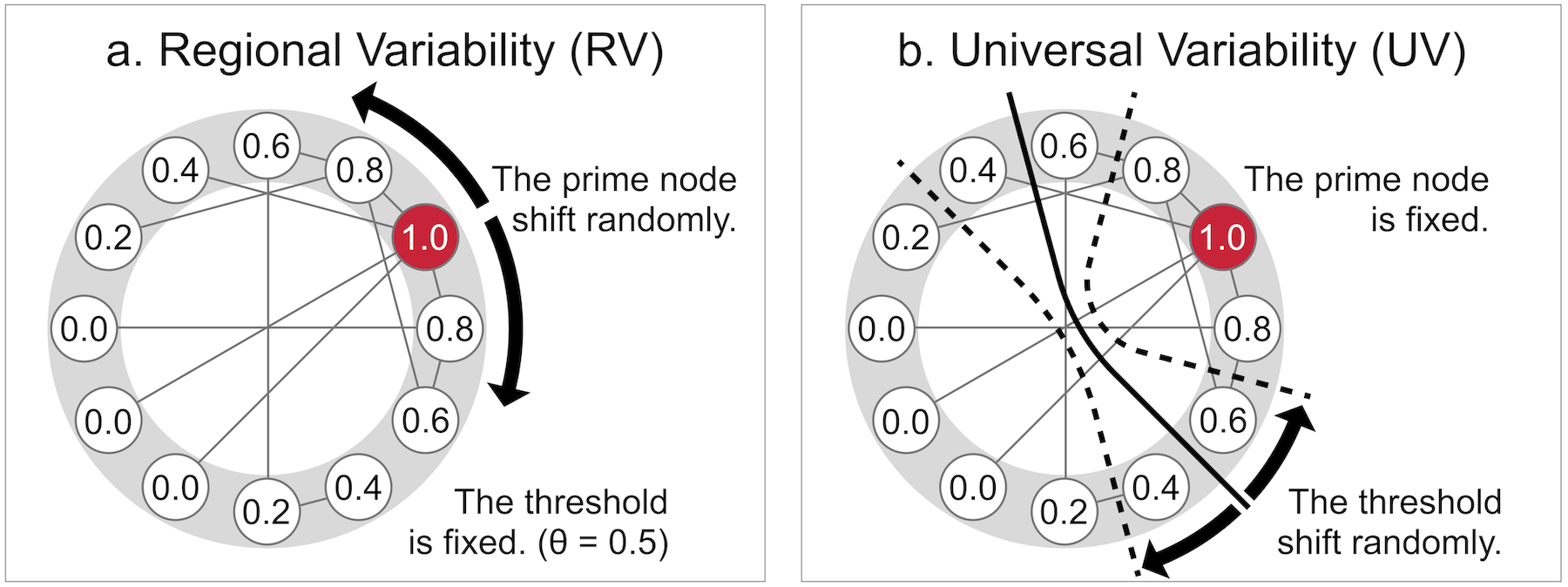}
    \caption{
Relationships, geographical structure, and EV.
Each small circle within the gray circle represents a group (agent, node), with the number inside indicating the resource value.
The gray ring represents the geographical structure, and each line connecting agents denotes a relationship (edge).
(a) In the RV model, the prime node shifts randomly within the geographical structure, and resources are then allocated to other nodes.
Following this, the game and reformation processes occur.
The resource threshold for reformation is fixed at $\theta = 0.5$.
(b) In the UV model, the prime node remains fixed, but the threshold $\theta$ fluctuates randomly, following an AR(1) process.
    }
    \label{fig:model}
\end{figure}

The geographic structure is modeled as a line segment with periodic boundary conditions, represented visually as a circle (Fig \ref{fig:model}).
$N$ agents are evenly spaced along the circle.
Although a one-dimensional spatial structure is used for simplicity, more realistic structures can be explored as empirical studies progress.
Mobility is not considered in this study to keep the model simple as an initial approach.
We plan to incorporate mobility in future work, as it is reasonable to assume that a group might move to a resource-rich region when their own resources become scarce.
However, the assumption of no mobility is not entirely unrealistic, as some studies have highlighted the tendency for settlement during the MSA \cite{Marean2010, Wadley2011, Kandel2012}.

The interaction structure defines the relationships between agents, which are not limited by the geographic arrangement.
These relationships affect the frequency of games (described later) and are subject to rewiring through reformations (also described later).
The network, where agents are represented as nodes and relationships as edges, is characterized as an undirected, unweighted, and dynamic graph.
The initial network is a regular network with a degree of $k = 4$.

\subsection*{EV}

The resource represents the amount of goods or wealth necessary for survival that a group (agent) can obtain from the natural environment (e.g., food, materials for stone tools required to gather food).
Resources are allocated to each agent at each time step, with the amount varying across different regions.
The node with the highest resource allocation is referred to as the prime node.
The further an agent is located from the prime node geographically, the less resource it receives, as determined by the resource decrement factor, $f_{RD}$.
Specifically, the resource allocated to agent $i$ is calculated as $r_i = r_p - |i - p| f_{RD}$.
Here, $r_i$ represents the resource allocated to agent $i$, $p$ is the index of the prime node, and $|i - p|$ represents the distance between $i$ and $p$, accounting for the boundary conditions, rather than the usual absolute value.
Additionally, there is a universal resource threshold, $\theta$, for all agents; any agent falling below this threshold must reformulate its strategy and relationships.

In our study, EV to refers to resource variability, represented by stochastic models.
EV can be divided into variability in distribution of resources among regions and in total quantity of resources across all regions.
Although variations in resource types are also an important consideration, we simplify the model assuming a single resource type.
The two forms of variability are termed regional variability (RV) and universal variability (UV).
RV is a type of EV in which the distribution of resource-rich and resource-poor regions changes over time; specifically, the prime node moves randomly.
The prime node’s index $p_t$ at time $t$ fluctuates according to a stochastic process expressed as $p_{t+1} = ( p_t + \Delta_t ) \mod N$.
Here, $\Delta_t$ is a random integer step uniformly distributed within the range $[ -\sigma_R, \sigma_R ]$, where $0 \leq \sigma_R \leq N / 2$.
UV represents another form of EV in which the total resource quantity fluctuates randomly over time, while the resource distribution between regions remains fixed.
This variability reflects large-scale fluctuations in resource availability across a wide area encompassing all regions.
However, to simplify the implementation, we fix the resource values and instead model UV by varying the threshold value $\theta$, which determines when reformation occurs.
The fluctuation is modeled by an AR(1) process \cite{Hasselmann1976, Vyushin2012, Salcedo2022}, $\theta_{t+1} = \mu_\theta (1 - \beta) + \theta_t \beta + \epsilon$, where $\mu_\theta$ is the expected value of $\theta$, $\beta$ is the autoregressive coefficient ($0 \leq \beta \leq 1$), and $\epsilon$ is a normally distributed noise term with mean 0 and standard deviation (SD) $\sigma_\theta$.
The intensity of RV is determined by the shift range of the prime node ($\sigma_R$), while the intensity of UV is influenced by the autoregressive coefficient ($\beta$) and the SD of the noise term ($\sigma_\theta$).
We examine the impact of EV on the evolution of cooperation under three scenarios: RV, UV, and combined variability (CV), where both the prime node shifts and the threshold $\theta$ fluctuates.

\subsection*{Game}

Communication over resources (such as primitive bartering, giving, looting) between agents is represented by simple pairwise games.
These games can only be played with an opponent who is connected through a network edge.
The probability that agent $i$ selects agent $j$ as its game opponent from its neighbors is $p^G_{i,j} = \frac{1}{n}$;
$n$ is the number of neighbors of $i$, and neighbors refer to an agent directly connected to $i$ in the interaction structure network.
The game procedure follows a pairwise public goods game (PGG) \cite{Hardin1968, Binmore1994, Kollock1998, Santos2008} with resource threshold considerations.
First, assume that agent $i$ selects agent $j$.
If the agent is C, it contributes a surplus resource $M_i = \max(r_i - \theta, 0)$.
If the agent is D, it does not contribute any resource.
The contributed resources are multiplied by a factor $b$ ($1 \leq b \leq 2$) and then equally divided between $i$ and $j$.
The payoff table is as follows:
\begin{equation}
\begin{array}{|c|c|c|}
\hline
  & C & D \\
\hline
C & R_i, R_j & S_i, T_j \\
\hline
D & T_i, S_j & P_i, P_j \\
\hline
\end{array}
\end{equation}
\begin{align}
R_i &= \left(\frac{b}{2} - 1\right) M_i + \frac{b}{2} M_j \\
S_i &= \left(\frac{b}{2} - 1\right) M_i \\
T_i &= \frac{b}{2} M_j \\
P_i &= 0
\end{align}

The social optimum, which maximizes the sum of both payoffs, is CC under $b > 1$.
The Nash equilibrium is when DD for $b < 2$.
Thus, a social dilemma exists across the defined range of $b$, except at the boundaries.
Additionally, by solving $T_i > R_i > P_i > S_i$, the condition for the game to qualify as a prisoner’s dilemma is $b > 1 + \frac{M_i - M_j}{M_i + M_j}$.
If this condition is not met, then $T_i > P_i > R_i > S_i$.

\subsection*{Reformation}

If, as a result of the games, an agent’s resource falls below the threshold $\theta$, it is considered to have failed to adapt to the environment, triggering a reformation of its strategy and network connections.
An agent $i$ that falls below $\theta$ randomly selects a role model.
The probability that $j$ is chosen as its role model is $p^R_j = \frac{r_j}{\sum_{k \in [1, .., N]} r_k}$.
The $i$ imitates $j$’s strategy, with mutation occurring at a probability $\mu$.
Additionally, agents that fall below the threshold disconnect all of their current relationships.
They then randomly select the same number of new neighbors as the number of disconnections and establish new connections.
The probability that agent $j$ is chosen as a new neighbor is proportional to $p^R_j$.
In other words, the higher the resource, the more likely an agent is to be chosen as a role model and a new neighbor.

\subsection*{Evaluation}

The simulation runs for $10,000$ generations, with $100$ independent simulations conducted for each parameter set (Table \ref{tab:parameters}).
Resource allocation and interactions, including games and reformations, occur once per generation.
The proportion of agents employing strategy C in each generation is referred to as the cooperation rate.
The average cooperation rate is calculated as the mean of the cooperation rates across trials, considering only the last $50\%$ of the generations.
To assess the effect of EV on the evolution of cooperation we compared the average cooperation rates across different parameters controlling RV and UV, along with other factors.

The effect of EV was assessed by comparing the average cooperation rates across various parameters.

\begin{table}[ht]
    \renewcommand{\arraystretch}{1.3}
    \centering
    \begin{tabular}{|c|l|c|}
        \hline
        Parameter & \multicolumn{1}{c|}{Description} & Value \\
        \hline
        $N$ & Number of agents & 100 \\
        $n^C_0$ & Initial number of C agents & 0 \\
        $k_0$ & Initial degree of the interaction structure network & 4 \\
        $t_{max}$ & Number of time steps (generations) for a simulation & 10,000 \\
        $trials$ & Number of simulations per parameter set & 100 \\
        \hline
        $r_{max}$ & Max resource & $1.0$ \\
        $f_{RD}$ & Resource decrement factor & $0.02$ \\
        $\sigma_R$ & Shift range of the prime node & $0$ to $49$ ($1$ steps)  \\
        $\mu_\theta$ & Expected value of threshold $\theta$ & $0.5$ \\
        $\beta$ & Autoregressive coefficient of the UV & $0.0$ to $0.9$ ($0.1$ steps) \\
        $\sigma_\theta$ & SD of the noise term of UV & $\{0.0, 0.1\}$ \\
        \hline
        $b$ & Multiplication factor for PGG & $1.0$ to $2.0$ ($0.1$ steps) \\
        $\mu$ & Mutation rate for strategy imitation & 0.01 \\
        \hline
    \end{tabular}
    \caption{Parameters}
    \label{tab:parameters}
\end{table}

\section*{Results}
\subsection*{Effect of RV}

We first examined the impact of RV in isolation, without considering UV.
Specifically, in each generation, the prime node randomly shifts within the range of $[-\sigma_R, \sigma_R]$, accounting for the periodic boundary condition.
The threshold $\theta$, which universally affects the resource welfare of all agents, is fixed at $0.5$.

\begin{figure}[ht]
    \centering
    \includegraphics[width=0.95\linewidth]{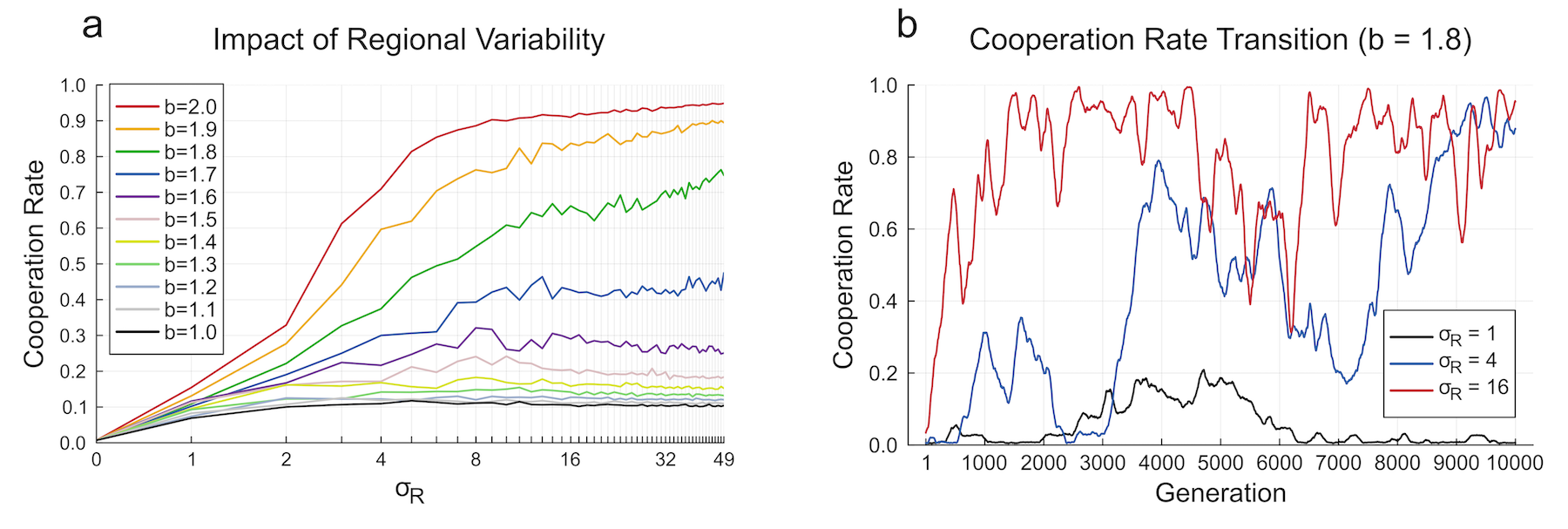}
    \caption{
The effect of RV.
(a) The effect of RV across $b \in [1.0, ..., 2.0]$.
The horizontal axis represents the intensity of RV, $\sigma_R$,
and the vertical axis shows the mean cooperation rate over the last $5,000$ generations, averaged across $100$ trials.
(b) Examples of cooperation rate transitions when $b = 1.8$ and $\sigma_R \in [1, 4, 16]$.
The horizontal axis represents generations, and the vertical axis shows the cooperation rate.
Higher RV tends to result in fluctuations of the cooperation rate at higher levels, though convergence is not observed.
    }
    \label{Fig:Regional}
\end{figure}

The results (Fig \ref{Fig:Regional}a) suggest that RV can promote the evolution of cooperation.
To elaborate, when the prime node is fixed ($\sigma_R = 0$), cooperation does not evolve.
As variability slightly increases ($\sigma_R = 1$), the cooperation rate rises to around $10\%$.
For $b \leq 1.7$, further increases in variability do not promote additional cooperation, although the cooperation rate remains higher than when $\sigma_R = 0$ or $1$.
However, for $b \ge 1.8$, greater variability further enhances cooperation.

Despite the overall positive effect of higher variability on cooperation, cooperation is not completely stable and fluctuates with temporary environmental changes.
For example, when $b = 1.8$ and $\sigma_R \in [1, 4, 16]$, the temporal transition (Fig \ref{Fig:Regional}b) shows that the cooperation rate rises and falls dramatically.

\subsection*{Effect of UV}

Next, we examined the effects of UV while excluding RV.
As previously defined, UV refers to fluctuations in the threshold $\theta$ across generations, which uniformly affects all agents.
The intensity of this variability is controlled by the autoregressive coefficient $\beta$ and the SD $\sigma_\theta$ of the noise term in the AR(1) stochastic model.
With the prime node fixed at $p = 1$, resources are allocated less as the distance from this node increases.

\begin{figure}[ht]
    \centering
    \includegraphics[width=0.95\linewidth]{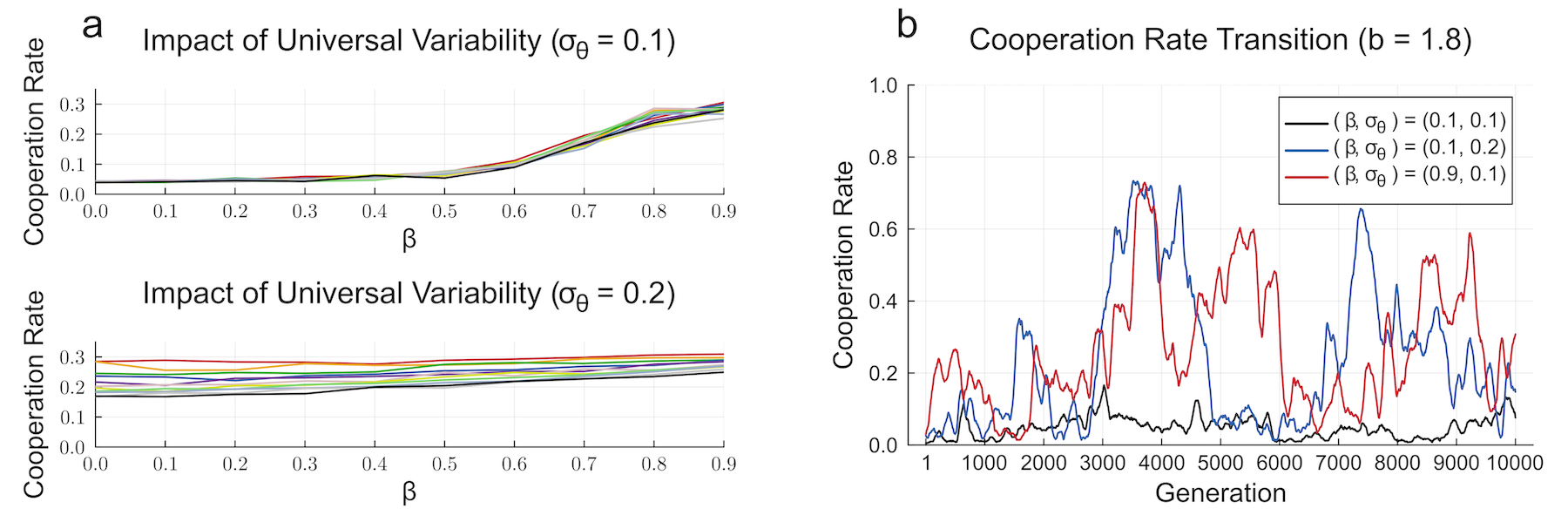}
    \caption{
The effect of UV.
(a) The effect of UV across $b \in [1.0, ..., 2.0]$ and $\sigma_\theta \in [0.1, 0.2]$.
The horizontal axis represents the intensity of UV, $\beta$,
and the vertical axis shows the mean cooperation rate over the last $5,000$ generations across $100$ trials.
(The line color scheme of these lines is same as in Fig \ref{Fig:Regional}a.)
(b) Examples of the cooperation rate transition when $b = 1.8$ and $(\beta, \sigma_\theta) \in [(0.1, 0.1), (0.1, 0.2), (0.9, 0.1)]$.
The horizontal axis represents generations, and the vertical axis shows the cooperation rate.
Higher UV tends to result in the cooperation rate fluctuating at higher levels, but convergence is not observed.
    }
    \label{Fig:Universal}
\end{figure}

We found that while UV promotes the evolution of cooperation to some extent, its effect is considerably more limited than that of RV.
For $\sigma_\theta = 0.1$, the cooperation rate gradually increases when $\beta$ exceeds 0.5, but it peaks at just $30\%$ (Fig \ref{Fig:Universal}a).
For $\sigma_\theta = 0.2$, the cooperation rate remains between $20\%$ and $30\%$, regardless of $\beta$, and increasing variability further does not affect these outcomes (Fig \ref{Fig:Universal}a).

\subsection*{Effect of CV}

\begin{figure}[ht]
    \centering
    \includegraphics[width=0.95\linewidth]{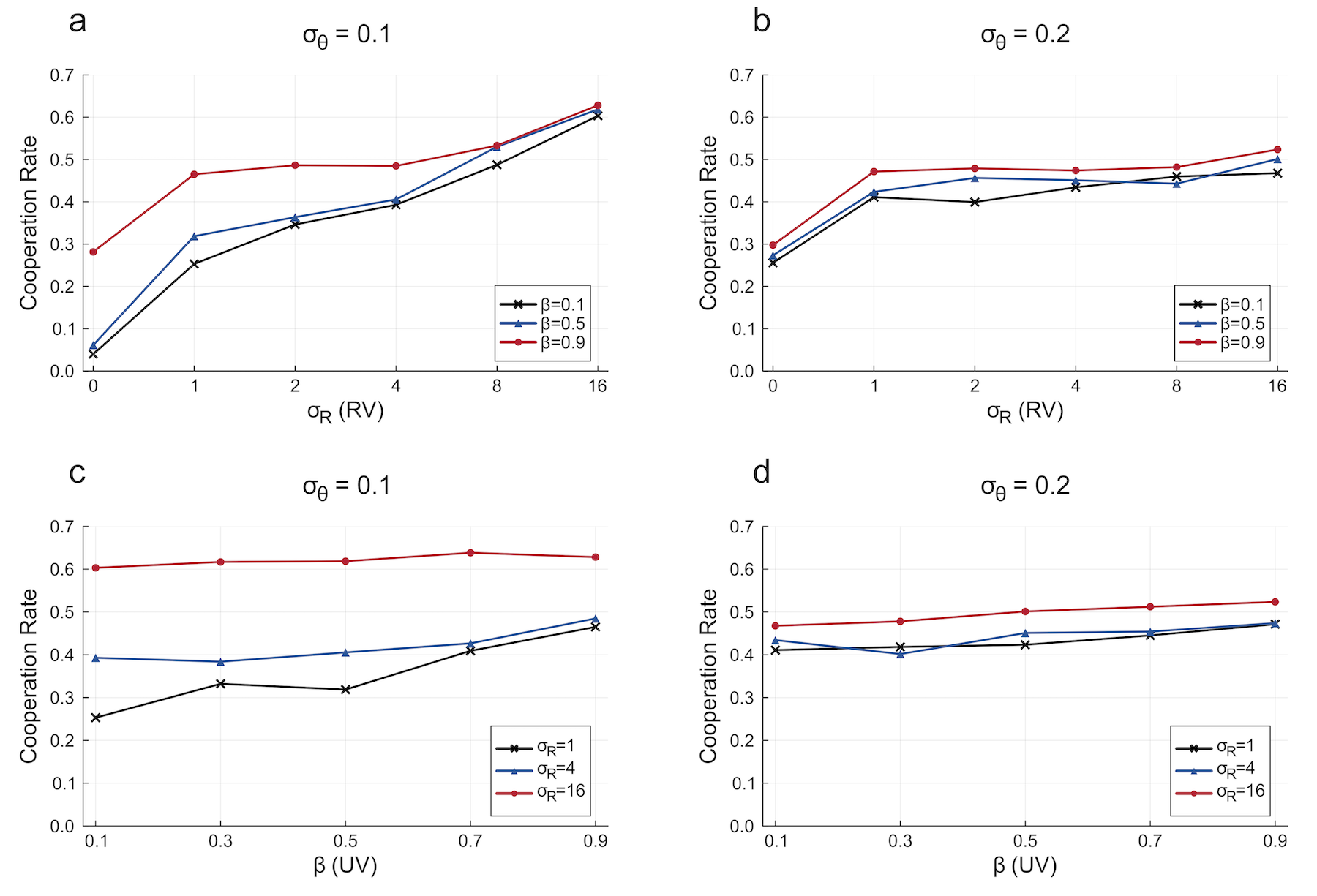}
    \caption{
The effect of CV ($b = 1.8$).
(a) and (b) represent cases with $\sigma_\theta = 0.1$, each with a different x-axis.
These plots show that RV increases the cooperation rate significantly, while UV has a limited effect.
(c) and (d) represent cases with $\sigma_\theta = 0.2$, with different x-axis.
The lines are almost flat except in the range $\sigma_R = 0$ to $1$,
indicating that when UV is too high, not only UV but even RV has no effect on the cooperation rate.
    }
    \label{fig:combination}
\end{figure}

We then examined the combined effects of both RV and UV.
The results (Fig \ref{fig:combination}) show that, consistent with the separate analyses above, RV strongly promotes the evolution of cooperation, while UV has much subtle effect.

\subsection*{Primary drivers of the results}

The results are driven by two key factors:
1. the effects of mutation and EV, which promote fluctuations in cooperation rates, and
2. the coevolution  of cooperation and network structure.

\begin{figure}[ht]
    \centering
    \includegraphics[width=1.0\linewidth]{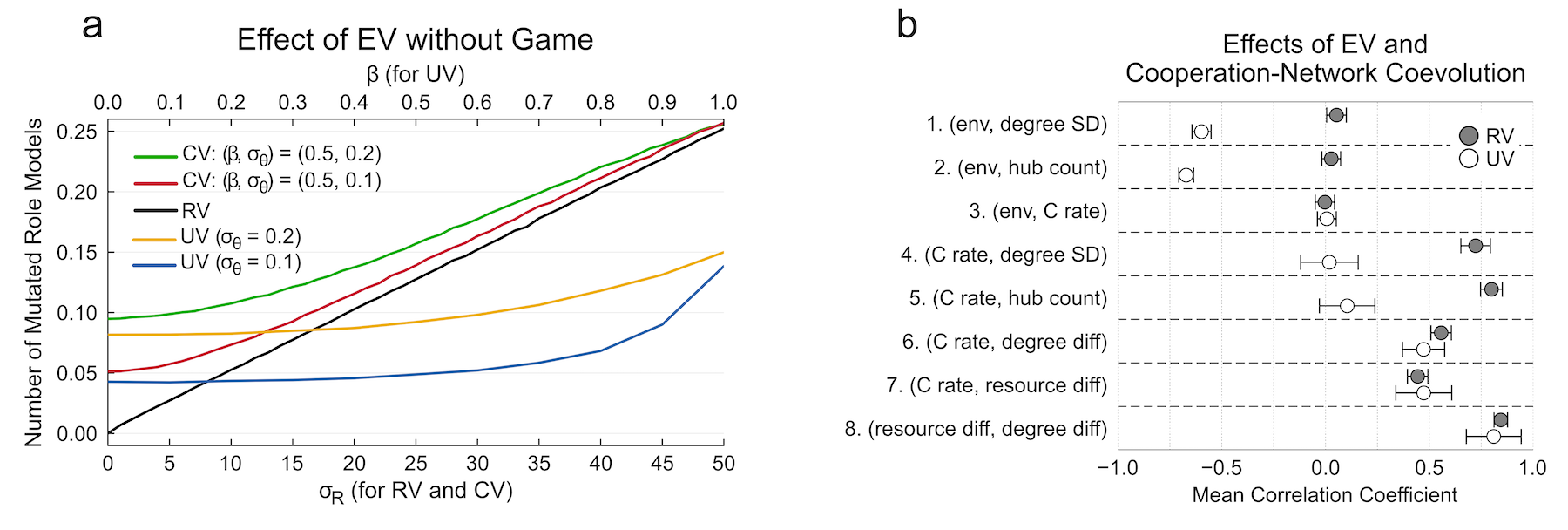}
    \caption{
Effects of EV on Cooperation.
(a) The number of mutated role models as a function of RV and UV.
The simulations are based on a model that excludes games.
Therefore, the figure shows how EV and reformation impact the system without the effect of games or network structure.
(b) Correlation analysis through time series ($10,000$ generations) between variables related to EV
(For RV, env refers to the shift distance of the prime node per generation; for UV, env refers to the value of threshold $\theta$),
network structure (degree SD: standard deviation of degrees, hub count: number of nodes with a degree of $10$ or more), and
cooperation (C rate: frequency of C, resource diff: the difference in average resources between C and D, and degree diff: the difference in average degree between C and D).
The mean correlation coefficients are averaged over $100$ trials for both RV and UV.
$b = 1.8$ for all simulations, $\sigma_R = 16$ for RV, and $\beta = 0.7$, $\sigma_{\theta} = 0.1$ for UV.
}
    \label{fig:mechanism}
\end{figure}

First, RV increases the fluctuations in the cooperation rate, rather than the rate itself.
Fig \ref{fig:mechanism}a shows the effect of EV on strategy distribution in a model that entirely excludes the effects of games and networks.
When these effects are excluded, changes in strategy distribution occur solely due to mutation and strategy updating.
The Y-axis represents the number of agents generated by mutation in one generation, who then serve as role models for strategy updating in the next generation.
These agents are the source of changes in strategy distribution.
The line for RV in the figure shows that as the variability increases, the number of mutated role models increases linearly.
Fig \ref{fig:mechanism}b ``3. (env, C rate)" shows that the time series of RV and the cooperation rate in each of the 100 trials are completely uncorrelated.
Furthermore, the results remain unchanged even when cross-correlation analysis is performed, accounting for time delays.
Therefore, it is evident that RV does not directly affect the cooperation rate but instead promote fluctuations in it.
This can be explained as follows: agents in poorer regions frequently undergo reformations and mutations.
When RV is small, agents rarely accumulate resources in the next generation, causing them to undergo reformations again.
However, when variability is large, agents may become resource-rich in the next generation, and they potentially survive to influence the strategy updates of other agents.
When RV is large, the cooperation rate is more likely to fluctuate up and down.

Notably, the analytical solution (\ref{eq:n_MR}) aligns closely with the simulation results for RV shown in Fig \ref{fig:mechanism}a.
\begin{align}
E[n_{MR}] &= \sum_{k=0}^n \mu^k (1 - \mu)^{n - k} k \frac{\sigma_R}{N} \notag \\
          &= \frac{n \mu}{N} \sigma_R \label{eq:n_MR}
\end{align}
In this equation, $E[n_{MR}]$ denotes the expected number of mutated role models, which are generated through mutation and later serve as role models for the strategy updates of other agents.
The right-hand side of the equation represents the expected number of mutated agents within a population of reformed agents, $n$, weighted by the effect of RV, $\sigma_R$, across the entire agent population, $N$.
This equation is simplified by applying the formula for the expectation of a binomial distribution.

Second, once the cooperation rate increases, it is easily sustained through the coevolution of cooperation and the network structure.
Fig \ref{fig:mechanism}b 4-8 shows the correlation between the cooperation rate (C rate), the heterogeneity of the network degrees (degree SD: standard deviation of degrees, hub count: number of nodes with a degree of 10 or more), the difference in the average resources between C and D, and the difference in the average degree between C and D.
These factors are strongly correlated in the cases of RV.
This can occur through the following process: as the cooperation rate increases, the average resources of C rise due to mutual support.
As the average resource of C increases, C acquires more connections than D.
This increases the heterogeneity of the network’s degree distribution.
As demonstrated in several studies \cite{Santos2005, Santos2006, Santos2008}, networks with heterogeneous degree distribution tend to promote cooperation.
In this way, as the cooperation rate increases, it enhances the heterogeneity of the network, which, in turn, further boosts the cooperation rate.
This creates a positive feedback loop that helps sustain the elevated cooperation rate once it has risen by chance.

In contrast, UV does not significantly promote cooperation for two reasons:
it fails to generate sufficient fluctuation in the cooperation rate, and
it inhibits the coevolution of cooperation and the network structure.
As shown by the two UV lines in Fig \ref{fig:mechanism}a, increasing variability does not substantially increase the number of mutated role models, unlike the linear relationship seen with RV.
Furthermore, when UV is intense and the threshold $\theta$ becomes very large, almost all agents undergo reformation.
This leads to a reduction in network heterogeneity, which is critical for the coevolution of cooperation and network structure.
This is reflected by the strong inverse correlation between UV and network heterogeneity (Fig \ref{fig:mechanism}b 1-2).
The factors that promote cooperation in RV are ineffective in the context of UV, which explains why UV does not significantly promote cooperation.
When RV and UV are combined in the CV model, the results remain consistent, RV promotes cooperation, while UV has a much smaller effect.

In summary, the observed patterns in cooperation rates can be attributed to the combined effects of mutation and EV, as well as the coevolution of cooperation and network structure.
Specifically, when these two factors work well together, as in the RV model, environmental change promotes cooperation.
However, when the first factor is weak and inhibits the second, as in the UV model, cooperation is less likely to evolve.

\section*{Discussion}

Building on Potts's VSH \cite{Potts1996, Potts1998}, we explored the effects of EV through two simplified models: RV and UV models.
Our results show that RV clearly promotes the evolution of cooperation, while UV has a much smaller effect.
The reason RV fosters cooperation is that it disrupts the distribution of strategies, and the cooperative states that occasionally emerge from these disturbances are sustained by the coevolution of cooperation and network structure.
In contrast, UV does not significantly promote cooperation because the disturbances it causes are insufficient and because it hinders the evolution of the network structure.
Even when both RV and UV are combined, only RV proves effective, with UV having little impact.
Thus, when EV's disturbance effect and the coevolution of cooperation and network structure align, EV facilitates the evolution of cooperation.
However, EV does not promote cooperation if it fails to create sufficient disturbance in strategy distribution or if it undermines the network’s heterogeneous structure.

Our study offers a new perspective on the evolution of cooperation and has the potential to guide future research on modern human behavior in archaeological contexts.
First, although some research on the evolution of cooperation considering EV has been conducted in biology and physics \cite{Brockhurst2007, Miller2015, Gokhale2016, Stojkoski2021, Assaf2013}, few studies have approached it from an anthropological perspective.
To address this gap, we offer a novel explanation for how EV may influence cooperative behavior, using a model that simulates interactions among human groups during the MSA.
Notably, we also identify a key condition for EV to promote cooperation: cooperation and network structure must coevolve effectively.
Second, our findings suggest that the VSH, which has primarily been linked to the CBH, may also be connected to the SBH, thus expanding the scope of VSH.
The conventional CBH explanation \cite{Potts1996, Potts1998, Schuck-Paim2008, Sol2008, Sol2009} is that the brains of the ``versatilists," evolved to rapidly adapt to new environments under EV, contributing to the development of modern human behavior.
In contrast, our novel interpretation, linking EV to SBH, suggests the following sequence.
EV promotes intergroup cooperation, as demonstrated in our study.
The cooperation likely leads to the emergence of more complex, interdependent societies, which introduce new social challenges such as communication, coordination, and conflict resolution across group boundaries.
The increasing complexities of these societies exerts selection pressure on cognitive abilities, favoring individuals who are better equipped to navigate these dynamics.
Over time, this process likely drives the evolution of modern human behavior, characterized by advanced intellectual abilities and complex social structures.
Finally, not previously mentioned, our network model, which incorporates both relationship and geographical structures, can be viewed as a type of multiplex network \cite{Wang2015}.
This perspective opens new avenues for mutual enrichment by enabling comparisons between our findings and the extensive body of research on multiplex and multilayer networks \cite{Gomez-Gardenes2012, Wang2015, Su2022a, Inaba2023}, potentially providing deeper insights into the dynamics of cooperation within complex social and spatial frameworks.

Although this study provides valuable insights into the relationship between EV and cooperation, several limitations should be considered for a more comprehensive understanding.
These limitations include the validity of our model in relation to real-world scenarios during the MSA, the extent to which our results explain actuality, and the need for further mathematical analysis.
We have prioritized simplicity in constructing the model, it is essential to verify which aspects align with reality and which do not.
This includes evaluating the EV patterns, geographical structure, population and grouping dynamics, the specifies of intergroup interactions (game), and the mechanisms through which collective behaviors propagate (reformation), as influenced by empirical research.
Additionally, population changes and mobility must be considered, and we plan to address mobility in future work.
We also need to wait for future empirical studies to assess how well our results correspond to the actual events of the MSA.
Furthermore, mathematical analysis is required to better understand the conditions under which EV and other dynamic structures can foster cooperation.
Although the complexity of the phenomenon makes it challenging to propose analytical solutions, theoretical research using simpler models should complement our findings.
Thus, while our study does not aim to establish universal laws, it presents a valuable hypothesis that is expected to guide future theoretical and empirical research in this area.

\section*{Supporting information}

\subsection*{Data and code availability}

All data, along with the code needed to run the simulations and generate the plots, are available at https://github.com/mas178/Inaba2024.

\section*{Contributions}
M.I. was responsible for designing and implementing the research, analyzing the results, and writing the manuscript.
E.A. supervised the research.
All authors reviewed and approved the final manuscript.

\nolinenumbers

\bibliography{bibliography}
\end{document}